\documentclass[conference]{IEEEtran}
\ifCLASSINFOpdf
\else
\fi
\usepackage[cmex10]{amsmath}
\usepackage{url}
\usepackage[T1]{fontenc}

\usepackage{multirow}
\usepackage{epsfig,endnotes}
\usepackage{cite}
\usepackage{pbox}
\usepackage{graphicx}

\newenvironment{packed_item}{
\begin{itemize}
  \setlength{\itemsep}{1pt}
  \setlength{\parskip}{0pt}
  \setlength{\parsep}{0pt}
}{\end{itemize}}

\begin{document}
\title{The Feasibility of Dynamically Granted Permissions:\\Aligning Mobile Privacy with User Preferences}

\author{
{\rm Primal Wijesekera$^{1}$, Arjun Baokar$^2$, Lynn Tsai$^2$, Joel Reardon$^2$, }\\
{\rm Serge Egelman$^2$, David Wagner$^2$, and Konstantin Beznosov$^1$}\\
$^1$University of British Columbia, Vancouver, Canada,\\
%\vspace{0.1em}
\{primal,beznosov\}@ece.ubc.ca\\
$^2$University of California, Berkeley, Berkeley, USA,\\
\{arjunbaokar,lynntsai,joel.reardon\}@berkeley.edu, \{egelman,daw\}@cs.berkeley.edu\\
% copy the following lines to add more authors
% \and
% {\rm Name}\\
%Name Institution
} % end author
\maketitle

\begin{abstract}
%\boldmath
Current smartphone operating systems regulate application permissions by prompting users on an ask-on-first-use basis. Prior research has shown that this method is ineffective because it fails to account for context: the circumstances under which an application first requests access to data may be vastly different than the circumstances under which it subsequently requests access. We performed a longitudinal 131-person field study to analyze the contextuality behind user privacy decisions to regulate access to sensitive resources. We built a classifier to make privacy decisions on the user's behalf by detecting when context has changed and, when necessary, inferring privacy preferences based on the user's past decisions and behavior. Our goal is to automatically grant appropriate resource requests without further user intervention, deny inappropriate requests, and only prompt the user when the system is uncertain of the user's preferences. We show that our approach can accurately predict users' privacy decisions 96.8\% of the time, which is a four-fold reduction in error rate compared to current systems.
\end{abstract}


% \fi
%%\IEEEpeerreviewmaketitle

\section{Introduction}
\label{sec:introduction}

One of the roles of a mobile application platform is to help users
avoid unexpected or unwanted use of their personal data~\cite{Enck2010}.
Mobile platforms currently use permission systems to regulate access to
sensitive resources, relying on user prompts to
determine whether a third-party application should be granted or denied
access to data and resources.
One critical caveat in this approach, however, is that mobile platforms
seek the consent of the user the first time a given application attempts to access a certain data type and then enforce the user's decision
for all subsequent cases, regardless of the circumstances surrounding
each access. For example, a user may grant an application access to location data because she is using location-based features, but by doing this, the application can subsequently access location data for behavioral advertising, which may violate the user's preferences.

Earlier versions of Android (5.1 and below) asked users to make privacy
decisions during application installation as an all-or-nothing ultimatum (ask-on-install):
either all requested permissions are approved or the application is not installed.
Previous research showed that
few people read the requested permissions at install-time and even fewer correctly
understood them~\cite{Felt2012}. 
Furthermore, install-time permissions do not present users with the context in which those permission will be exercised,
which may cause users to make suboptimal decisions not aligned with their
actual preferences. For example, Egelman et al.~observed that when an application requests access to location data without providing context, users are just as likely to see this as a signal for desirable location-based features as they are an invasion of privacy~\cite{Egelman2012}. Asking users to make permission decisions at runtime---at
the moment when the permission will actually be used by the application---provides more context (i.e., what they were doing at the time that data was requested)~\cite{Felt2012b}.
However, due to the high frequency of permission
requests, it is not feasible to prompt the user every time data is accessed~\cite{wijesekera2015}.

In iOS and Android M, the user is now prompted at runtime the first time an application
attempts to access one of a set of ``dangerous'' permission types (e.g.,
location, contacts, etc.). This \emph{ask-on-first-use} (AOFU) model is
an improvement over ask-on-install (AOI). Prompting users the first time an application uses one of the designated
permissions gives users a better sense of context: their knowledge of what they
were doing when the application first tried to access the data
should help them determine whether the request is appropriate.
However, Wijesekera et al.~showed that AOFU fails to meet user expectations
over half the time, because it does not account for the varying contexts of
future requests~\cite{wijesekera2015}.

The notion of \emph{contextual integrity} suggests that many
permission models fail to protect user privacy because they fail to
account for the context surrounding data flows~\cite{Nissenbaum2004}. That is, privacy violations occur when sensitive resources are used in ways that defy
users' expectations. We posit that more effective permission models must focus
on whether resource accesses are likely to defy users' expectations in a given
context---not simply whether the application was authorized to
receive data the first time it asked for it. Thus, the challenge for system
designers is to correctly infer when the context surrounding a
data request has changed, and whether the new context is likely to be deemed
``appropriate'' or ``inappropriate'' for the given user. Dynamically regulating data access based on the context requires more 
user involvement to understand users' contextual preferences. If users are asked to make privacy decisions
too frequently, or under circumstances that are seen as low-risk, they may
become habituated to future, more serious, privacy decisions.
On the other hand, if users are asked
to make too few privacy decisions, they may find that the system has acted
against their wishes. Thus, our goal is to automatically determine {\it when} and under {\it what} circumstances to present users with runtime prompts.

%Over-privileged
%applications also distract users from actual permissions they
%should be focusing on and contribute to user habituation towards
%future permission requests~\cite{Felt2011b}.

%A
%significant portion of applications are over-privileged, which means that they
%request access to data types that they do not actually need to function. This is
%primarily due to lack of awareness by developers~\cite{Felt2011b}, and it results in user
%habituation towards future permission requests.

To this end, we collected real-world Android usage data in order to explore
whether we could infer users' future privacy decisions based on their past
privacy decisions, contextual circumstances surrounding applications' data requests, and users'
behavioral traits.
We conducted a field study where 131 participants used Android phones
that were instrumented to gather data over an average of 32 days per participant.
Also, their phones periodically prompted them to make privacy decisions
when applications used sensitive permissions,
and we logged their decisions.
Overall, participants wanted to block 60\% of these requests.
We found that AOFU yields 84\% accuracy, i.e., its policy
agrees with participants' prompted responses 84\% of the time.
AOI achieves only 25\% accuracy.

We designed new techniques that use machine learning
to automatically predict how users would respond to prompts, so that
we can avoid prompting them in most cases, thereby reducing user burden.
Our classifier uses the user's past decisions in related situations
to predict their response to a particular permission prompt.
The classifier outputs a prediction and a confidence score; if the classifier
is sufficiently confident, we use its prediction, otherwise we prompt the
user for their decision.
We also incorporate information about the user's behavior in other security
and privacy situations to make inferences about their preferences: whether they have a screen lock activated,
how often they visit HTTPS websites, and so on.
%One of the novel contributions of this work is our finding that this
%behavioral information is predictive of user privacy preferences; furthermore, it requires no user involvement to be helpful in solving this problem.
We show that our scheme achieves 95\% accuracy (a $4 \times$ reduction in
error rate over AOFU) with significantly less user involvement than the {\it status quo}.

%better
%understand the contextuality behind users privacy decisions and to train a
%classifier to automatically make permission decisions on the user's behalf by
%inferring user preferences and context.

%Although, AOFU was proposed as an improvement over AOI, it requires more
%active user involvement (i.e., responding to runtime prompts). Previous research has
%established, however, that too much {\it unnecessary} user involvement leads to habituation~\cite{Felt2012,Felt2012b}. To this end, we built a classifier that uses passively observed behavioral traits to limit the number of prompts that users see. In this manner, we were able to yield a
%three-fold improvement over AOI with absolutely no user involvement. This is a significant
%step towards reducing unnecessary user involvement in the learning phase. A classifier
%which uses runtime, contextual, and past behavior was able to reduce the misprediction error
%by 45\% over AOFU while reducing the user involvement by 59\% compared AOFU;
%this is a near four-fold improvement over AOI.

The specific contributions of our work are the following:
\begin{packed_item}
\item We conducted the first known large-scale study on quantifying the effectiveness of ask-on-first-use permissions.
\item We show that a significant portion of the studied participants make
contextual decisions on permissions using the foreground application and the
visibility of the permission-requesting application.
\item We show how a machine-learned model can incorporate context and
better predict users' privacy decisions.
\item To our knowledge, we are the first to use passively observed traits
to infer future privacy decisions on a case-by-case basis at runtime.
\end{packed_item}

\section{Related Work}

There is a large body of work demonstrating that install-time prompts fail
because users do not understand or pay attention to them~\cite{Kelley2012,
Gorla2014, Wei2012}. When using install-time prompts,
users often do not understand which
permission types correspond to which sensitive resources and are surprised by
the ability of background applications to collect information~\cite{Felt2012,
Thompson2013, Jung2012}. Applications also transmit a large amount of location
or other sensitive data to third parties without user consent~\cite{Enck2010}.
When possible risks associated with these requests are revealed to users, their
concerns range from annoyance to wanting to seek retribution~\cite{Felt2012c}.

To mitigate some of these problems, systems have been developed to track
information flows across the Android system~\cite{Enck2010, Gibler2012,
Klieber2014} or introduce finer-grained permission control into
Android~\cite{Hornyack2011, Almohri2014, Shebaro2014}, but
many of these solutions increase user involvement significantly, which can
lead to habituation. Additionally, many of these proposals are useful only to
the most-motivated or technically savvy users. For example, many such
systems require users to configure complicated control panels, which many
are unlikely to do~\cite{yee05}. Other approaches involve static analysis
in order to better understand how applications {\it could} request
information~\cite{Au2012, Bodden2013, Felt2011b}, but these say little about
how applications {\it actually} use information. Dynamic analysis improves upon
this by allowing users to see how often this information is requested in
real time~\cite{Spreitzenbarth2013, wijesekera2015, Enck2010}, but substantial
work is likely needed to present that information to average users in a
meaningful way. Solutions that require runtime prompts (or other user
interruptions) need to also minimize user intervention, in order to prevent
habituation.

Other researchers have developed recommendation systems to recommend
applications based on users' privacy preferences~\cite{Zhu2014}. Systems have
also been developed to predict what users would share on mobile social
networks~\cite{Bilogrevic2013}, which suggests that future systems could
potentially infer what information users would be willing to share with
third-party applications. By requiring users to self-report privacy
preferences, clustering algorithms have been used to define user privacy
profiles even in the face of diverse preferences~\cite{sadeh2014}. However,
researchers have found that the order in which information is requested has an
impact on prediction accuracy~\cite{wu2014improving}, which could mean that
such systems are only likely to be accurate when they examine actual user
behavior over time (rather than relying on one-time self-reports).

Liu et al.\ clustered users by privacy preferences and used ML
techniques to predict whether to allow or deny an application's request for
sensitive user data~\cite{Liu2014}. However, their dataset was collected from a
set of highly privacy-conscious individuals---those choosing to install a
permission-control mechanism. Furthermore, the researchers removed
``conflicting'' user decisions, in which a user chose to deny a permission for
an application, and then later chose to allow it. However, these conflicting decisions
happen nearly 50\% of the time in the real world~\cite{wijesekera2015}, and
accurately reflect the nuances of user privacy preferences; they are not
experimental mistakes, and therefore models need to account for them. In fact, previous work found that users commonly
reassess privacy preferences after usage~\cite{almuhimedi2015}. Liu et al.\
also expect users to make 10\% of permission decisions manually, which, based
on field study results from Wijesekera et al., would result in being prompted
every three minutes~\cite{wijesekera2015}. This is obviously impractical. Our
goal is to design a system that can automatically make decisions on behalf of
users, that accurately models their preferences, while also not over-burdening
them with repeated requests.

Closely related to this work, Liu et al.~\cite{liu2016} performed a field
study to measure the effectiveness of a Privacy Assistant that offers recommendations
to users on privacy settings that they could adopt based on each user's privacy
profile---the privacy assistant predicts what the user might want based on the inferred privacy profile and static analysis of the third-party application.
While this approach increased user awareness on resource usage, the recommendations
are static: they do not consider each application's access to sensitive data on a case-by-case basis. Such a coarse-grained approach goes against previous work suggesting that people do want to vary their decisions based on contextual circumstances~\cite{wijesekera2015}.
A blanket approval or denial of a permission to a given application carries a considerable risk of
privacy violations or loss of desired functionality. In contrast, our work tries to infer the appropriateness of a
given request by considering the surrounding contextual cues and how the user has behaved
in similar situations in the past, in order to make decisions on a case-by-case basis using dynamic analysis. Their dataset was collected from a
set of highly privacy-conscious and considerably tech-savvy individuals, which might
limit the generalization of their claims and findings, whereas we conducted a field study on a more representative sample.

Nissenbaum's theory of contextual integrity suggests that permission models
should focus on information flows that are likely to defy user
expectations~\cite{Nissenbaum2004}. There are three main components involved in
deciding the appropriateness of a flow~\cite{Barth2006}: the context in which
the resource request is made, the role played by the agent requesting the
resource (i.e., the role played by the application under the current context), and
the type of resource being accessed. Neither previous nor currently deployed
permission models take all three factors into account. This model could be used to
improve permission models by automatically granting access to data when the
system determines that it is appropriate, denying access when it is
inappropriate, and prompting the user only when a decision cannot be made
automatically, thereby reducing user burden.

\emph{Access Control Gadgets} (ACGs) were proposed as a mechanism to tie sensitive resource
access to certain UI elements~\cite{roesner2012, roesner2013, ringer2016}. Authors
posit that such an approach will increase user expectations since a significant
portion of participants expected a UI interaction before a sensitive resource
usage, giving users an implicit mechanism to control access and increasing the
awareness on resource usage. The biggest caveat in this approach is tying a UI
interaction to each sensitive resource access is practically impossible due to the
high frequency at which these resources are accessed~\cite{wijesekera2015}, and due to the fact that
many legitimate resource accesses occur without user initiation~\cite{Felt2012b}.

Wijesekera et al.\ performed a
field study~\cite{wijesekera2015} to operationalize the notion of ``context,''
so that an operating system can differentiate between appropriate and
inappropriate data requests by a single application for a single data type.
They found that users' decisions to allow a
permission request were significantly correlated with that application's
visibility: in this case, the context is using or {\it not} using the requesting application.
They posit visibility of the application could be a strong
contextual cue that influences users' responses to permission prompts.
%This suggests that using the name of the requesting application,
%the visibility of that application, and
%the specific data type requested as a limited kind of context
%for deciding whether or not an information
%flow is likely to be appropriate. 
They also observed that privacy
decisions were highly nuanced, and therefore a one-size-fits-all model is
unlikely to be sufficient; a given information flow may be
deemed appropriate by one user and inappropriate by another user. They
recommended applying machine learning in order to infer individual users'
privacy preferences.

To achieve this, research is needed to determine what factors affect user privacy decisions and how to use those factors to make privacy decisions on the user's behalf.
While we cannot automatically capture everything involved in
Nissenbaum's notion of context, we can try for the next-best thing:
we can try to detect when context has likely changed (insofar as to decide whether or not a different privacy decision should be made for the same application and data type),
by seeing whether the circumstances surrounding a data request are similar to previous requests or not.

\section{Methodology}
%How we executed the analysis.

\begin{table}[t] %t?
\small
\begin{center}
\begin{tabular}{|l|p{4.3cm}|}
\hline
\textbf{Permission Type} & \textbf{Activity} \\ \hline

\begin{tabular}[c]{@{}l@{}}\textsc{access\_wifi\_state}\end{tabular} & View nearby SSIDs \\ \hline
\textsc{nfc} & Communicate via NFC \\ \hline
\begin{tabular}[c]{@{}l@{}}\textsc{read\_history\_bookmarks}\end{tabular} & Read users' browser history \\ \hline
\begin{tabular}[c]{@{}l@{}}\textsc{access\_fine\_location}\end{tabular} & Read GPS location \\ \hline
\begin{tabular}[c]{@{}l@{}}\textsc{access\_coarse\_location}\end{tabular} & \begin{tabular}[c]{@{}l@{}}Read network-inferred location\\ (i.e., cell tower and/or WiFi)\end{tabular} \\ \hline
\begin{tabular}[c]{@{}l@{}}\textsc{location\_hardware}\end{tabular} & Directly access GPS data \\ \hline
\textsc{read\_call\_log} & Read call history \\ \hline
\textsc{add\_voicemail} & Read call history \\ \hline
\textsc{read\_sms} & Read sent/received/draft SMS \\ \hline
\textsc{send\_sms} & Send SMS \\ \hline
\textsc{*internet} & Access Internet when roaming \\ \hline
\begin{tabular}[c]{@{}l@{}}\textsc{*write\_sync\_settings}\end{tabular} &
\begin{tabular}[c]{@{}l@{}}Change application sync \\ settings when roaming\end{tabular} \\ \hline
\end{tabular}
\end{center}
\begin{flushleft}
\caption{Felt et al.\ proposed granting a select set of 12 permissions at runtime so that users have contextual information to infer why the data might be needed~\cite{Felt2012b}. Our instrumentation omits the last two permission types
(\textsc{internet} \& \textsc{write\_sync\_settings}) and records information about the other 10.}
\label{tbl:perm_list}
\end{flushleft}
\end{table}

We collected data from 131 participants to understand what factors could be used to infer whether a permission request is likely to be deemed appropriate by the user.

Previous work by Felt et al.\ made the argument that certain permissions are
appropriate for runtime prompts, because they protect sensitive resources---and
therefore require user intervention---and because viewing the prompt at runtime imparts
additional contextual information about why an application might need
the permission~\cite{Felt2012b}. Similarly, Thompson et al.\ showed that other permission requests could be replaced with audit mechanisms, because they represent either reversible changes or are low enough risk to not warrant habituating the user to prompts~\cite{Thompson2013}.
We collected information about 10 of the 12 permissions Felt et al.\ suggest
are best-suited for runtime prompts;
we omitted \textsc{internet} and \textsc{write\_sync\_settings},
   since we did not expect any participant to be roaming while using our
   instrumentation,
and focused on the remaining 10 permission types (Table
									  \ref{tbl:perm_list}).
While there are many other sensitive permissions beyond this set, Felt et al.\ concluded that the others are best handled by other mechanisms (e.g., install-time prompts, ACGs, etc.).

\begin{figure}[t]
\centering
\includegraphics[scale=0.29]{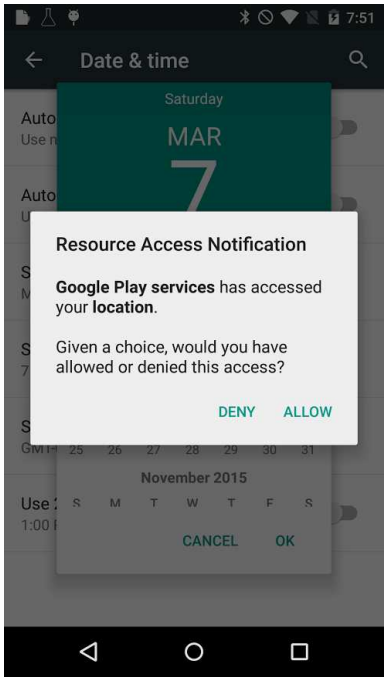}
\begin{flushleft}
\caption{A screenshot of an ESM prompt.}
\label{fig_scrnshot}
\end{flushleft}
\end{figure}

We used the Experience Sampling Method (ESM) to collect ground truth data about users' privacy preferences~\cite{esm86}. ESM involves repeatedly questioning participants {\it in situ} about a recently observed event; in this case, we probabilistically asked them about an application's recent access to data on their phone, and whether they would have permitted it, if they had been given the choice. We treated participants' responses to these ESM probes as our main dependent variable (Figure \ref{fig_scrnshot}).

We also instrumented participants' smartphones to obtain data about their privacy-related behaviors and the frequency with which applications accessed protected resources. The instrumentation required a set of modifications to the Android operating system and flashing a custom Android version onto participants' devices. To facilitate such experiments, the University of Buffalo offers non-affiliated academic researchers access to the PhoneLab panel~\cite{phonelab}, which consists of more than 200 participants. All of these participants had LG Nexus 5 phones running Android 5.1.1 and the phones were periodically updated over-the-air (OTA) with custom modifications to the Android operating system. Participants can decide when to install the OTA update, which marks their entry into new experiments. During our experiment period, different participants installed the OTA update with our instrumentation at different times, thus we neither have data on all PhoneLab participants, nor for the entire period. Our OTA update was available to participants for a period of six weeks, between February 2016 and March 2016. At the end of the study period, we emailed participants a link to an exit survey to collect demographic information. Our study was approved by the relevant institutional review board (IRB).

%(2016-02-05 to 2016-03-17) - phonelab study period.
%We extended the instrumentation available in PhoneLab to collect required data from users.

\subsection{Instrumentation}
\label{sec:instrumentation}

The goal of our instrumentation was to collect as much runtime and behavioral data as could be observed from the Android platform, with minimal impact on performance. We collected three categories of data: behavioral information, runtime information, and user decisions. We made no modifications to any third-party application code, which means that our dynamic analysis techniques could be used on any third-party Android application.

%To log the data, we relied on the logging framework implemented by PhoneLab, which was built on top of the \textit{logcat} framework. 

\begin{table}[t]
\small
\centering
\begin{tabular}{|c|l|}
\hline
\textbf{Type} & \textbf{Event Recorded} \\ \hline
\multirow{21}{*}{\begin{tabular}[c]{@{}c@{}}Behavioral \\ Instrumentation\end{tabular}} & Changing developer options \\ \cline{2-2} 
 & Opening/Closing security settings \\ \cline{2-2} 
 & Changing security settings \\ \cline{2-2} 
 & Enabling/Disabling NFC \\ \cline{2-2} 
 & Changing location mode \\ \cline{2-2} 
 & Opening/Closing location settings \\ \cline{2-2} 
 & Changing screen-lock type \\ \cline{2-2} 
 & Use of two factor authentication\\ \cline{2-2} 
 & Log initial settings information \\ \cline{2-2} 
 & User locks the screen \\ \cline{2-2} 
 & Screen times out \\ \cline{2-2} 
 & App locks the screen \\ \cline{2-2} 
 & Audio mode changed \\ \cline{2-2} 
 & Enabling/Disabling speakerphone \\ \cline{2-2} 
 & Connecting/Disconnecting headphones \\ \cline{2-2} 
 & Muting the phone \\ \cline{2-2} 
 & Taking an audio call \\ \cline{2-2} 
 & Taking a picture (selfie vs.\ non-selfie) \\ \cline{2-2} 
 & Visiting an HTTPS link in Chrome \\ \cline{2-2} 
 & Responding to a notification \\ \cline{2-2} 
 & Unlocking the phone \\ \hline
\multirow{2}{*}{\begin{tabular}[c]{@{}c@{}}Runtime\\ Information\end{tabular}} & An application changing the visibility \\ \cline{2-2} 
 & Platform switches to a new activity \\ \hline
\multirow{2}{*}{\begin{tabular}[c]{@{}c@{}}Permission\\ Requests\end{tabular}} & An app requests a sensitive permission \\ \cline{2-2} 
 & ESM prompt for a selected permission \\ \hline
\end{tabular}
\begin{flushleft}
\caption{Instrumented events that form our feature set}
\label{tbl:events}
\end{flushleft}
\end{table}

Table \ref{tbl:events} contains the complete list of behavioral and runtime events our
instrumentation recorded. The behavioral data fell under several categories, all
chosen based on several hypotheses that we had about the types of behaviors that
might correlate with privacy preferences: web browsing behavior, screen locking
behavior, third party application usage behavior, audio preferences, call
habits, camera usage patterns (selfie vs.\ non-selfie), and behavior related to security settings. For example, we hypothesized that someone who manually locks their device screen (as opposed to letting it time out) might be more privacy-conscious than someone who takes many speakerphone calls or selfies.

%The primary purpose of recording user behaviors are to observe how much time a user voluntarily spends on making security and privacy related decisions by changing default settings or re-visiting decisions users have made earlier (security settings, location settings), observing behaviors that could be implicitly indicative of their privacy preferences such as choices they have made on screen locks, how careful are they with their web browsing habits such as how often do they use Chrome incognito tabs, how often they get warnings from visiting suspicious websites. We also have collected other observable traits that could be indicative of privacy consciousness such as how often they take pictures, their audio preferences, how active they are with audio calls using the phone, etc. 

We also collected runtime information about the context of each permission request, including the visibility of the requesting application at the time of request (i.e., whether it was running in the foreground or not) and what the user was doing when the request was made (i.e., the name of the foreground application). The visibility of an application reflects the extent to which the user was likely aware that the application was running; if the application was in the foreground, the user had cues that the application was running, but if it was in the background, then the user was likely not aware that the application was running and therefore might find the permission request unexpected.
We also collected information about which Android \texttt{Activity} was active in the application;\footnote{An Android \texttt{Activity} represents the application screen and UI elements currently exposed to the user.} depending on the design of the application, this might tell us only that the user was browsing with Firefox or might provide fine-grained information such as differentiating between reading a news feed vs.\ searching for a user's profile on Facebook. We monitored processes' memory priority levels to determine the visibility of all active processes.

We recorded every time that an application used one of the 10 permissions
mentioned earlier.
We also recorded the exact Android API invoked by a third-party application to
determine precisely what information was requested.

Finally, once per day we randomly selected one of these permission requests
and prompted the user about them at runtime (Figure \ref{fig_scrnshot}).
We used weighted reservoir sampling to select a permission request
to prompt about. We weight permissions based on their
frequency of occurrence seen by the instrumentation; the most-frequent
permission request has a higher probability of being shown to participants using
ESM\@. We prompted participants a maximum of three times for each unique
combination of requesting application, permission, and visibility of the
requesting application (i.e., background vs.\ foreground). We tuned the wording of the prompt to make it clear that the request had just occurred and their response would not affect the system (a deny response would not actually deny data). These responses serve as the ground truth for all the analysis mentioned in the remainder of the paper.

The intuition behind using weighted reservoir sampling is to focus more
on the frequently occurring permission requests over rare ones. Common
permission requests contribute most to user
habituation due their high frequency. Thus, it is more important to learn
about user privacy decisions on highly frequent permission requests over the 
rare ones, which might not risk user habituation or annoyance (and the context of rare requests may be less likely to change).
%In \S\ref{sec:steady_state} we present an approach based on the decision confidence
%which we can use to learn more about rare and unseen events and then start inferring
%users' decisions on them. Thus we believe the sampling method does not 
%affect any of the results presented in the remainder of the paper.

\subsection{Exit Survey}

At the end of our data collection period, PhoneLab staff emailed participants a link to our online exit survey, which they were incentivized to complete with a raffle for two \$100 Amazon gift cards. The survey gathered demographic information and qualitative information on their privacy preferences. Of the 203 participants in our experiment, 53 fully completed the survey, and another 14 partially completed it. Of the 53 participants to fully complete the survey, 21 were male, 31 were female, and 1 undisclosed. Participants ranged from 20 to 72 years of age (\begin{math}\mu \end{math} = 40.83, \begin{math} \sigma \end{math}= 14.32). Participants identified themselves as 39.3\% staff, 32.1\% students, 19.6\% faculty, and 9\% other. Only 21\% of the survey respondents had an academic qualification in STEM, which suggests that the sample is unlikely to be biased towards tech-savvy users.

\subsection{Summary}
%data summarizing.
We collected data from February 5 to March 17, 2016. PhoneLab allows any participant to opt-out of an experiment at any time. Thus, of the 203 participants who installed our custom Android build, there were 131 who used it for more than 20 days. During the study period, we collected 176M events across all participants (31K events per participant/day). Our dataset consists of 1,686 unique applications and 13K unique activities. Participants also responded to 4,636 prompts during the study period. We logged 96M sensitive permission requests, which translates to roughly one sensitive permission request every 6 seconds per participant. For the remainder of the paper, we only consider the data from the 131 participants who used the system for at least 20 days, which corresponds to 4,224 ESM prompts.

Of the 4,224 prompts, 55.3\% were in response to ACCESS \_WIFI\_STATE, when trying to access Wifi SSID information that could be used to infer the location of the smartphone; 21.0\%, 17.3\%, 5.08\%, 0.78\%, and 0.54\%  were from accessing location directly, reading SMS, sending SMS, reading call logs, and accessing browser history, respectively. A total of 137 unique applications triggered prompts during the study period. %143 unique applications ran in the foreground when the participants were prompted. 
Of the 4,224 prompts, participants wanted to deny 60.01\% of them, and 57.65\% of the prompts were shown when the requesting application was running in the foreground or the user had visual cues that the application was running (e.g., notifications). A Wilcoxon signed rank test with continuity correction revealed a statistically significant difference in participants' desire to allow or deny a permission request based on the visibility of the requesting application ($p<0.0152$, $r=0.221$), which corroborates previous findings~\cite{wijesekera2015}.

\section{Types of Users} 
\label{impact-visibility}

We hypothesized that there may be different types of users based on how
they want to disclose their private information with third parties. It is
imperative to identify these different sub-populations since
different permission models affect users differently based on their
privacy preferences; performance numbers averaged across a user population could be
misleading since different sub-populations might react differently to
the same permission model.

While our study size was too small to effectively apply
clustering techniques to generate classes of users, we 
did find a meaningful distinction using the denial
rate (i.e., the percentage of prompts to which users wanted to deny access). 
We aggregated users by their denial rate in 10\% increments and examined how these different participants considered the surrounding
contextual circumstances in their decisions.

We discovered that application visibility was a significant factor for users with a denial rate of 10--90\%, but not for users
with a denial rate of 0--10\% or 90--100\%.
We call the former group \emph{Contextuals}, as they care about the
surrounding context (i.e., they make nuanced decisions), and the latter group \emph{Defaulters}, because, as we now
show, they don't seem to take the surrounding context into account when
they make privacy decisions.
%they tend to either allow application permissions or deny them and did not vary their decision-making based on circumstances.
%This seems to adhere to
%\cite{wijesekera2015}, which posited that the visibility of the application
%requesting the permission is a strong contextual cue used by users when deciding
%whether to allow or deny a permission request.

\begin{figure}[t]
\centering
\includegraphics[scale=0.38]{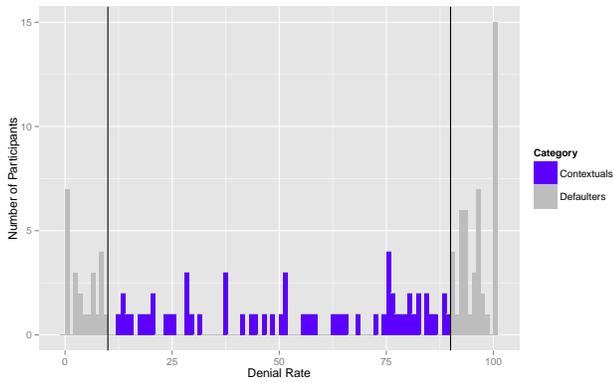}
\begin{flushleft}
\caption{Histogram of users based on their denial rate. \textit{Defaulters} tended to allow or deny almost all requests without regard for contextual cues, whereas \textit{Contextuals} considered the visibility of the requesting application.}
\label{fig_denial_ratio}
\end{flushleft}
\end{figure}

\textit{Defaulters} accounted for 53\% of 131
participants and \textit{Contextuals} accounted for 47\%. A Wilcoxon
signed-rank test with continuity correction revealed a statistically
significant difference in \textit{Contextuals'} responses based on requesting
application visibility ($p<0.013$, $r=0.312$), while for \textit{Defaulters} there
was no statistically significant difference ($p=0.227$, $r=0.215$). That is, \textit{Contextuals} 
used visibility as a contextual cue, when deciding the appropriateness of a given permission 
request, whereas \textit{Defaulters} did not vary their decisions 
based on this cue. %and instead consistently chose one option for the duration of the experiment.
Figure \ref{fig_denial_ratio} shows the distribution of users based on their
denial rate. Vertical lines indicate the borders between \emph{Contextuals}
(light gray) and \emph{Defaulters} (dark gray). Observe that \emph{Defaulters} appear at
both ends of the denial-rate spectrum, while \emph{Contextuals} fully occupy the
space between them.

 In the remainder of the paper, we use our
\emph{Contextuals--Defaulters} 
categorization to measure how current and proposed models affect
these two sub-populations, issues unique to these sub-populations, and ways to
address these issues.

\begin{table}[t]
\small
\centering
\scalebox{.9}{
\begin{tabular}{|l|r|r|r|r|}
\hline
\textbf{Policy}  & \textbf{Contextuals} & \textbf{Defaulters} & \textbf{Overall} & \textbf{Prompts} \\ \hline
AOI        & 44.11\%            & 6.00\%      & 25.00\%      & 0.00               \\ \hline
\textbf{AOFU-AP} & \textbf{64.49\%}       & \textbf{93.33\%}      & \textbf{84.61\%}   & \textbf{12.34}   \\ \hline
AOFU-APV           & 64.28\%                  & 92.85\%               & 83.33\%             & 15.79              \\ \hline
AOFU-A$_{F}$PV    & 66.67\%                  & 98.95\%               & 84.61\%             & 16.91              \\ \hline
AOFU-VP            & 58.65\%                 & 94.44\%               & 78.04\%              & 6.43               \\ \hline
AOFU-VA            & 63.39\%                  & 93.75\%               & 84.21\%             & 12.24              \\ \hline
AOFU-A             & 64.27\%                  & 93.54\%               & 83.33\%             & 9.06               \\ \hline
AOFU-P             & 57.95\%                  & 95.45\%               & 82.14\%           & 3.84               \\ \hline
AOFU-V             & 52.27\%                  & 95.34\%               & 81.48\%             & 2.00               \\ \hline
\end{tabular}
}
\begin{flushleft}
\caption{The accuracy and number of different possible ask-on-first-use combinations. A: Application requesting the permission, P: Permission type requested, V: Visibility of the application requesting the permission, A$_{F}$: Application running in the foreground when the request is made. AOFU-AP is the policy used in Android Marshmallow i.e., asking (prompting) the user for each unique application, permission combination. The table also differentiates policy numbers based on the subpopulation of \textit{Contextuals}, \textit{Defaulters}, and across all users.
}
\label{tbl:aofu_perf}
\end{flushleft}
\end{table}

\section{Ask-On-First-Use Permissions} 
\label{state-of-the-art}
Ask-on-first-use (AOFU) is the current Android permission model, which was first adopted in Android 6.0
(Marshmallow). AOFU works by prompting the user whenever an application requests
a \emph{dangerous} permission for the first time; the user's response to this
prompt is thereafter applied whenever the same application requests the same permission.
As of November 2016, only 24.3\% of
Android users have Android Marshmallow or a higher version~\cite{gdistribution}, and of those, those who have
upgraded from a previous version only see runtime permission prompts for freshly-installed
applications.

For the remaining 75.7\% of users, the system policy is
ask-on-install (AOI), which automatically allows all runtime permission requests. 
During the study period, all of our participants had AOI running as the default
permission model. Because all runtime permission requests are allowed in AOI, 
any of our ESM prompts that the user wanted to deny correspond to
mispredictions under the AOI model (i.e., the AOI model granted access to the data 
against users' actual preferences).  Table \ref{tbl:aofu_perf}
shows the expected median accuracy for AOI, as well as several other possible variants that we discuss in this section. The low median accuracy for \textit{Defaulters} was due
to the significant number of people who simply denied most of the prompts. The prompt
count is zero for AOI because it does not prompt the user during runtime; users are only
shown permission prompts at installation.

More users will have AOFU
in the future, as they upgrade to Android 6.0 and beyond. To the best
of our knowledge, no prior work has looked into quantifying the effectiveness of AOFU
systematically; this section presents analysis of AOFU based on prompt
responses collected from participants and creates a baseline against which to measure
our system's improvement.
We simulate how AOFU performs through our ESM prompt responses. Because AOFU is deterministic, each user's response
to the first prompt for each \textit{application:permission} combination tells us how the AOFU model would respond for subsequent requests by that same combination. For participants who responded to more than one prompt for each combination,
we can quantify how often AOFU would have been correct for subsequent requests. Similarly,
we also measure the accuracy for other possible policies that the platform could use to decide whether to prompt the user. For example, the status quo is for the platform to prompt the user for each new \textit{application:permission} combination, but how would accuracy (and the number of prompts shown) change if the policy were to prompt on all new combinations of \textit{application:permission:visibility}?
%Finally, over 95\% of actual users are currently using the conventional
%ask-on-install policy, and
%so we also analyze the effectiveness of the ask-on-install policy compared to AOFU\@.

Table \ref{tbl:aofu_perf} shows the expected median accuracy\footnote{The presented numbers---except for average prompt count, which was normally distributed---are median values, because the distributions were skewed.} for each policy based
on participants' responses. For each policy, \textit{A} represents the application
requesting the permission, \textit{P} represents the requested permission,
\textit{V} represents the visibility of the requesting application, and A$_{F}$
represents the application running in the foreground when a sensitive permission
request was made. For instance, AOFU-AP is the policy where the user will be
prompted for each new instance of an \textit{application:permission} combination, which is the Android 6.0 model. 
The last column shows the number of runtime prompts a participant would see under each
policy over the duration of the study, if that policy were to be implemented. 
Both AOFU-AP and AOFU-A$_F$PV show about a $4.9\times$ reduction in error rate compared to AOI;
AOFU-A$_F$PV would require more prompts over AOFU-AP, though yields a similar overall accuracy rate.\footnote{While AOFU-A$_F$PV has greater {\it median} accuracy when examining \textit{Defaulters} and \textit{Contextuals} separately, because the distributions are skewed, the median overall accuracy is identical to AOFU-AP when combining the groups.} Moving forward,
we focus our analysis only on AOFU-AP (i.e., the current standard).

Instances where the user wants to deny a permission and the policy instead allows it (false
positives) are \textit{privacy violations}, because they expose more information to the application than the user desires. Instances
where the user wants to allow a permission, but the policy denies it
(false negatives) are \textit{functionality losses}. This is because the
application is likely to lose some functionality that the user desired when it
is incorrectly denied a permission. Privacy violations and functionality losses were 
approximately evenly split between the two categories for AOFU-AP:
median privacy violations and median functionality losses were 6.6\% and 5.0\%, respectively.

%The combination of \textit{application:permission:visibility} has
%the lowest privacy violations, but also has a marginally higher number of runtime
%prompts over \textit{application:permission}. Marshmallow has a policy similar to
%\textit{application:permission} and it has the overall best accuracy over the rest of the
%other possible AOFU combinations. 

The AOFU policy works well for \textit{Defaulters},
because---by definition---they tend to be consistent after their initial responses for each
combination, which increases the accuracy of AOFU\@. In contrast, the decisions of
\textit{Contextuals} vary due to other factors beyond just the
application requesting the permission and the requested permission type. Hence,
the accuracy of AOFU for \textit{Contextuals} is significantly lower than the
accuracy for \textit{Defaulters}. This distinction shows that learning privacy
preferences for a significant portion of users requires a deeper understanding of
other factors affecting their decisions, such as behavioral tendencies and
contextual cues. As Table \ref{tbl:aofu_perf} suggests, superficially adding
more contextual variables (such as visibility of the requesting application)
does not necessarily help to increase the accuracy of the AOFU policy.

Our estimated accuracy numbers for AOFU may be inflated because AOFU in deployment (Android 6 and above) does not filter out permission requests that do not reveal any
sensitive information. For example, an application can request the
\textsc{access\_fine\_location} permission to check whether the phone has a 
specific location provider, which does not leak sensitive information (i.e., the system will check whether the requesting application has been granted the location permission, but no location data will actually be revealed to the application by this particular system call). Our AOFU simulation uses the invoked function to determine if sensitive data was {\it actually} accessed, and only prompts in those cases (in the interest of avoiding any false positives). Currently deployed AOFU in Marshmallow does not make this distinction. For example, Android users will see a permission request prompt when the application examines the list of location providers, and if the permission is granted, the user will not subsequently see prompts when location data is actually captured. Previous work showed that 79\% of first-time permission requests do not reveal any sensitive information~\cite{wijesekera2015}, and nearly 33.9\% of applications that request these sensitive permission types do not access sensitive data at all. The majority of AOFU prompts in Marshmallow are therefore effectively false positives, which incorrectly serve as the basis for future decisions. Given this, the average accuracy for AOFU is likely less than the numbers presented in Table \ref{tbl:aofu_perf}. We therefore consider our estimates of AOFU to be an upper bound.

\section{Learning Privacy Preferences} 
\label{sec:learning}

Table \ref{tbl:aofu_perf} shows that a significant portion of users (the 47\%
classified as \textit{Contextuals}) make privacy decisions that depend on
factors other than the application requesting the permission, the
permission requested, and the visibility of the requesting application.
To make decisions on behalf of the user, we must understand what other
factors affect their privacy decisions. We built a machine learning model
trained and tested on our labeled dataset of 4,224 prompts collected from 131
users over the period of 42 days. This approach is equivalent to training a
model based on runtime prompts from hundreds of users and using it to predict
those users' future decisions.
%This process also helped us determine which user behavior aspects relate strongly to user privacy decisions. 

We focus the scope of this work by making the following assumptions. We assume
that the platform, i.e., the Android OS, is trusted to manage and enforce
permissions for applications. We assume that applications must go through the
platform's permission system to gain access to protected resources.
We assume that we are in a non-adversarial machine-learning setting wherein the adversary does
not attempt to circumvent the machine-learned classifier by exploiting knowledge
of its decision-making process---though we do present a discussion of this problem and potential
solutions in Section~\ref{sec:disc}.

\subsection{Feature Selection}
\label{sec:feature_selection}

\begin{table}[t]
\centering
\begin{tabular}{|l|l|l|}
\hline
\textbf{\begin{tabular}[c]{@{}l@{}}Feature\\ Group\end{tabular}} & \textbf{Feature} & \textbf{Type} \\ \hline
\multirow{10}{*}{\begin{tabular}[c]{@{}l@{}}Behavioral\\ Features\\ (B)\end{tabular}} & \begin{tabular}[c]{@{}l@{}}Number of times a website is loaded to \\ the Chrome browser.\end{tabular} & Numerical \\ \cline{2-3} 
 & \begin{tabular}[c]{@{}l@{}}Out of all visited websites, the proportion\\ of HTTPS-secured websites.\end{tabular} & Numerical \\ \cline{2-3} 
 & The number of downloads through Chrome. & Numerical \\ \cline{2-3} 
 & \begin{tabular}[c]{@{}l@{}}Proportion of websites requested location\\ through Chrome.\end{tabular} & Numerical \\ \cline{2-3} 
 & \begin{tabular}[c]{@{}l@{}}Number of times PIN/Password was used to\\ unlock the screen.\end{tabular} & Numerical \\ \cline{2-3} 
 & Amount of time spent unlocking the screen. & Numerical \\ \cline{2-3} 
 & \begin{tabular}[c]{@{}l@{}}Proportion of times screen was timed out\\ instead of pressing the lock button.\end{tabular} & Numerical \\ \cline{2-3} 
 & Frequency of audio calls. & Numerical \\ \cline{2-3} 
 & Amount of time spent on audio calls. & Numerical \\ \cline{2-3} 
 & Proportion of time spent on silent mode. & Numerical \\ \hline
\multirow{4}{*}{\begin{tabular}[c]{@{}l@{}}Runtime\\ Features\\ (R1)\end{tabular}} & Application visibility (True/False) & Categorical \\ \cline{2-3} 
 & Permission type & Categorical \\ \cline{2-3} 
 & User ID & Categorical \\ \cline{2-3} 
 & Time of day of permission request & Numerical \\ \hline
\multirow{2}{*}{\begin{tabular}[c]{@{}l@{}}Aggregated\\ Features\\ (A)\end{tabular}} & \begin{tabular}[c]{@{}l@{}}Average denial rate for (A1)\\ application:permission:visibility\end{tabular} & Numerical \\ \cline{2-3} 
 & \begin{tabular}[c]{@{}l@{}}Average denial rate for (A2)\\ application$_F$:permission:visibility\end{tabular} & Numerical \\ \hline
\end{tabular}
\begin{flushleft}
\caption{The complete list of features used in the ML model evaluation. All the numerical values under behavioral group are normalized per day. We use one-hot encoding for categorical variables. We normalized numerical variables by making each one a z-score relative to its own average.}
\label{tbl:feature_list}
\end{flushleft}
\end{table}

Using the behavioral, contextual, and aggregate features shown
in Table \ref{tbl:events}, we constructed 16K candidate
features, formed by combinations of specific applications and actions.
Then, we selected 20 features by measuring Gini importance through random
   forests~\cite{Louppe2013}, significance testing for correlations, and
   singular value decomposition (SVD). SVD was particularly helpful to address the sparsity and high dimensionality issues caused by features generated based on application and activity usage.
Table \ref{tbl:feature_list} lists the 20 features used in the rest of this work.

The behavioral features (\textit{B}) that proved predictive relate to browsing 
habits, audio/call traits, and locking behavior. All behavioral features were 
normalized per day/user and were scaled in the actual model. Features relating 
to browsing habits included the number of websites visited, the proportion of 
HTTPS-secured links visited, the number of downloads, and proportion of sites 
visited that requested location access. Features relating to locking behavior 
included whether users employed a passcode/PIN/pattern, the frequency of screen 
unlocking, the proportion of times they allowed the screen to timeout instead of 
pressing the lock button, and the average amount of time spent unlocking the 
screen. Features under the audio and call category were the frequency of audio 
calls, the amount of time they spend on audio calls, and the proportion of time 
they spent on silent mode.

%Although there were few specific sets of activity usage were correlated with their privacy preferences, there were not predictive enough in the model due to their sparsity.
%such as number of websites visited daily, the proportion of those that were HTTPS-secured, how often they download files in Chrome, and how often they see location notification requests. Call and locking habits were indicative as well, including the number of calls made daily, amount of time spent with the phone in non-silent mode, time spent unlocking the phone (for those with PINs or passwords), and the proportion of times the user allowed to phone screen to timeout rather than actively locking it.

Our runtime features (\textit{R1}/\textit{R2}) include the requesting application's visibility, permission requested, and time of day a permission request occurred. Initially, we included the user ID to account for user-to-user variance, but as we discuss below, we subsequently removed it. Surprisingly, the name of the application requesting the permission was not predictive. Other features based on the requesting application, such as application popularity, similarly failed to be predictive. %A detailed analysis on this observation is given in \S\ref{sec:contextual-integrity}. 

%Initially, we tried three different types of learning models : Mixed Effect Logistic Regression, Random Forest, and Support Vector Machine. The introduction of random effects from the mixed effect model increased the accuracy significantly. Random effects capture within-user interactions (i.e., for same set of permission, application, visibility combination different users act differently). Random effects help the model to group the train set based on the user and treat data from each user differently from the rest. The large role of random effects also shows that users follow some consistent policy in their decision-making and that this policy is often unique for a user. 

Different users may have different ways of perceiving privacy threats posed by 
the same permission request. To account for this, the learning algorithm should 
be able to determine how each user treats permission requests in order to 
accurately predict their future decisions. To quantify the difference between 
users in how they perceive the threat posed by the same set of permission 
requests, we introduced a set of \textit{aggregate features} that could be 
measured at runtime and that might partly capture users' privacy preferences. We 
compute the average denial rate for each unique combination of 
\textit{application:permission:visibility} (\textit{A1}) and of 
\textit{permission:application$_{F}$\footnote{The application running in the 
foreground when the permission is requested by another application.}:visibility} 
(\textit{A2}). These aggregate features indicate how the user responded to 
previous prompts associated with that combination. As expected, after we 
introduced the aggregate features, the relative importance of the user ID 
variable diminished and so we removed it (i.e., users no longer needed to be 
uniquely identified). We define \textit{R2} as \textit{R1} without the user 
ID\@.

%to compare against the different AOFU combinations we propose in \S\ref{state-of-the-art}. This feature can be computed as a running average by the system after every prompt, and which works in practice, as we explore in \S\ref{online-model}.

\begin{table}[]
\centering
\small
\begin{tabular}{|l|r|r|r|}
\hline
\textbf{Feature Set} & \textbf{Contextuals} & \textbf{Defaulters} & \textbf{Overall} \\ \hline
%B                    & 67.48\%              & 96.00\%             & 83.21\%          \\ \hline
R1                   & 69.30\%              & 95.80\%             & 83.71\%          \\ \hline
R2 + B               & 69.48\%              & 95.92\%             & 83.93\%          \\ \hline
%R2 + A1              & 71.41\%              & 99.10\%             & 89.60\%          \\ \hline
%R2 + A2              & 70.83\%              & 98.08\%             & 90.10\%          \\ \hline
R2 + A        		 & 75.45\%              & 99.20\%             & 92.24\%          \\ \hline
\end{tabular}
\begin{flushleft}
\caption{The median accuracy of the machine learning model for different feature groups across different sub populations.}
\label{tbl:acc_feature}
\end{flushleft}
\end{table}

\subsection{Inference Based on Behavior}
\label{sec:behave_inference}

One of our main hypotheses is that passively observing users' behaviors could
help infer their future privacy decisions.
To this end, we instrumented Android to collect a wide array of behavioral data, listed in Table \ref{tbl:events}. We categorize our behavioral
instrumentation into interaction with Android privacy/security settings, locking behavior, audio settings and call habits, web browsing habits, and application usage habits. After the feature selection process (\S\ref{sec:feature_selection}), we found that only locking behavior, audio habits, and web browsing habits correlated with privacy behaviors. Please refer Appendix \ref{app:behaveinforgain} for more information on feature importance.

We trained an SVM model with an RBF kernel on only the behavioral and runtime features listed in Table~\ref{tbl:feature_list}, excluding user ID. The 5-fold cross validation accuracy (with random splitting) was 83\% across all users. This first setup assumes we have prior knowledge of previous privacy decisions to a certain extent from each user before inferring their future privacy decisions, so it is primarily relevant after the user has been using their phone for a while. However, the biggest advantage of using behavioral data is that it can be observed passively without any active user involvement (i.e., no prompting).

To measure the extent to which we can infer user privacy decisions with \textit{absolutely no user involvement} (and without any prior data on a user), we utilized leave-one-out cross validation. In this second setup, when a new user starts using a smartphone, we assume there
is a ML model which is already trained with behavioral data and privacy decisions collected from a selected set of other users. We then measured the efficacy of such a model to predict the privacy decisions of a new user, purely based on passively observed behavior, without prompting that new user at all. This is an even stricter lower bound on user involvement, which essentially mandates that a user has to make no effort to indicate privacy preferences, something that no system currently does.

We performed leave-one-out cross validation for each of our 131 participants, meaning we predicted a single user's privacy decisions using a model trained using the data from the other 130 users' privacy decisions and behavioral data. The only input for each test user was the passively observed
behavioral data and runtime data surrounding each request. The model yielded a median accuracy of 75\%, which is a 3X improvement over AOI. Furthermore, AOI requires users to make active decisions during the installation of an application, which our second model does not require.

Examining only behavioral data with leave-one-group-out cross validation yielded a median accuracy of 56\% for \textit{Contextuals}, while for \textit{Defaulters} it was 93.01\%. Although, prediction using solely behavioral data fell short of AOFU-AP for \textit{Contextuals}, 
it yielded a similar median accuracy for \textit{Defaulters}; AOFU-AP required 12 prompts to reach this level of accuracy, whereas our model would not have resulted in any prompts. This relative success presents the significant observation that behavioral features, observed passively without user involvement, are useful in learning user privacy preferences. This provides the potential to open entirely new avenues of user learning and reduce the risk of habituation.

\subsection{Inference Based on Contextual Cues} 
\label{sec:accuracy}

Our SVM model with a RBF kernel produced the best accuracy. The results in the remainder of this section are trained and tested with five-fold cross validation with random splitting for a SVM model with a RBF kernel using the \textit{ksvm} library in R\@. In all instances, the training set was bootstrapped with an equal number of allow and deny data points to avoid training a biased model. For each feature group, all hyperparameters were tuned through grid search to achieve highest accuracy. All the numerical values under the behavioral group were normalized per day. We used one-hot encoding for categorical variables. We normalized numerical variables by making each one a z-score relative to its own average. Table \ref{tbl:acc_feature} shows how the median accuracy changes with different feature groups. As a minor note, the addition of the mentioned behavioral features to runtime features performed only marginally better; this could be due to the fact that those two groups do not complement each other in predictions. In this setup, we assume that there is a single model across all the users of Android.

%The fact that RBF kernel outperforms the linear kernel in the SVM, proves that majority of user decisions aren't linear and involved many different factors.

%We believe this may be due to the fact that behavioral features are good at predicting overall privacy predispositions, but runtime features are far more important in determining answers to individual permission request prompts. This is supported by the fact that nearly all behavioral features statistically significantly correlate with a user's denial rate, as revealed by the Kendall rank correlation test.

By incorporating user involvement in the form of prompts, we can use our 
aggregate features to increase the accuracy for \textit{Contextuals}, slightly 
less so for \textit{Defaulters}. The aggregate features primarily capture how 
consistent users are for particular combinations (i.e., 
\textit{application:permission:visibility}, \textit{application:permission}, 
\textit{appli-cation$_{F}$:permission:visibility}), which greatly affects 
accuracy for \textit{Contextuals}. \textit{Defaulters} have high accuracy with 
just runtime features (\textit{R1}), as they are likely to stick with a default 
allow or deny policy regardless of the context surrounding a permission. Thus, 
even without any aggregate features (which do not impart any new information 
about this type of user), the model can predict privacy preferences of 
\textit{Defaulters} with a high degree of accuracy. On the other hand, 
\textit{Contextuals} are more likely to vary their decision for a given 
permission request. However, as the accuracy numbers in Table 
\ref{tbl:acc_feature} suggest, this variance is correlated with 
some contextual cues. The high predictive power of aggregate 
features indicates that they may be capturing the contextual cues used by 
\textit{Contextuals} to make decisions.

The fact that both \textit{application:permission:visibility} and 
\textit{$application_{F}$:permission:visibility} are highly predictive (Appendix \ref{app:infogain}) 
indicates that user responses for these combinations are consistent. 
The high consistency could relate to the notion that the visibility and the
foreground application (application$_F$\footnote{Even when the requesting application
is running visible to the user, the foreground application could still be different from
the requesting application since the only visible cue of the requesting application
could be a notification on notification bar.}) are strong contextual cues people 
use to make their privacy decisions; the only previously studied contextual 
cue was the visibility of the application requesting the sensitive 
data~\cite{wijesekera2015}. We offer a hypothesis for why foreground application 
could be significant: the sensitivity of the foreground application (i.e., 
high-sensitivity applications like banking, low-sensitivity applications like 
games) might impact how users perceive threats posed by requests. Irrespective 
of the application requesting the data, users may be likely to deny the request 
because of the elevated sense of risk. We discuss this further in 
\S\ref{sec:disc}.

The model trained on feature sets \textit{R2}, \textit{A1} and
\textit{A2} had the best accuracy (and fewest privacy violations). For the
remainder of the paper, we will refer to this model unless otherwise noted.
We now compare AOFU-AP (the status quo as of Android 6.0 and above, presented in Table
\ref{tbl:aofu_perf}) and our model (Table \ref{tbl:acc_feature}).
Across all users, our model reduced the error rate from
15.38\% to 7.76\%, which is nearly a two-fold
improvement. 

%While both approaches perform relatively
%well for \textit{Defaulters}, the ML model has a 4.72\%
%lower error rate. For \textit{Contextuals}, the ML
%model's improvements are much more dramatic, increasing
%accuracy from 64.49\% to 92.45\%. This gain is
%largely due to the contextual cues that the model
%takes into account (i.e., aggregate features). This shows that
%users do make contextual decisions rather than just
%basing their decision on application and permission, contrary to what
%AOFU assumes. That is, the aggregate features capture a
%notion of context, and these users' decisions are
%consistent across these notions of context. 

Mispredictions (errors) in the ML model were split
between privacy violations and functionality losses (54\% and 46\%).
Deciding which error type is more acceptable is
subjective and depends on factors like the usability issues surrounding
functionality losses and gravity of privacy violations. However, the (approximately) even split
between the two error types shows that the ML is not biased towards one
particular decision (denying vs.\ allowing a request).
Furthermore, the area under the ROC curve (AUC), a metric used to
measure the fairness of a classifier, is also
significantly better in the ML model (0.936 as opposed
to 0.796 for AOFU). This
indicates that the ML model is equally good at
predicting when to both allow and deny a permission
request, while AOFU tends to lean more towards one
decision. In particular, with the AOFU policy, users
would experience privacy violations for 10.01\% of
decisions, compared to just 4.2\% with the ML model.
Privacy violations are likely more costly to the user
than functionality loss, as denied data can always be
granted at a later time, but disclosed data cannot be
taken back.

% We can bias the ML model away from privacy violations simply by altering parameters; such an approach is not available with AOFU\@.

While increasing the number of prompts improves classifier accuracy, it
plateaus after reaching its maximum accuracy, at a point we call the \textit{steady state}. For some users, 
the classifier might not be able to infer their
privacy preference effectively, regardless of the number of prompts.  
As a metric to measure the effectiveness of the ML model, we measure the confidence of the model in the
decisions it makes, based on prediction class probabilities.\footnote{To calculate the class probabilities, we used the KSVM library in R. It employs a technique proposed by Platt et al.~\cite{lin2007note} to produce a numerical value for each class's probability.} In cases where the
confidence of the model is below a certain threshold, the system should use a
runtime prompt to ask the user to make an explicit decision. Thus, we looked
into the prevalence of low-confidence predictions among the current
predictions. With a 95\% confidence interval, on average across five folds,
low-confidence predictions accounted for less than 10\% of all predictions. The
remaining high-confidence predictions (90\% of all predictions) had an average
accuracy of 96.2\%, whereas predictions with low confidence were only predicted
with an average accuracy of 72\%. \S\ref{sec:steady_state} goes into this
aspect in detail and estimates the rate at which users will see prompts in
steady state. 

The caveat in our ML model is that AOFU-AP only resulted in 12 prompts on average per user during the study, while our model averaged 24. The increased prompting stems from multiple prompts for the same combination of \textit{application:permission:visibility}, whereas in AOFU, prompts are shown only once for each \textit{application:permission} combination. During the study period, users on average saw 2.28 prompts per unique combination. While multiple prompts per combination help the ML model to predict future decisions more accurately, it risks habituation, which may eventually reduce the reliability of the labeled data. 

The evaluation setup mentioned in the current section does not have a specific strategy to select the training set. It randomly splits the data set into the 5 folds and picks 4 out of 5 as the training set. In a real-world setup, the platform needs a strategy to carefully select the training set so that the platform can learn most of the user's privacy preferences with a minimum number of prompts. The next section presents an in-depth analysis on possible ways to reduce the number of prompts needed to train the ML model.

\section{Learning Strategy}
\label{simulation}

%This section presents different approaches of implementing a ML model in the real world. The analysis of the practicality of the ML model will be done in three phases: Bootstrapping, steady state, and online learning. The bootstrapping phase occurs when the ML model sees a new user and has no prior information on their privacy preferences. Steady state analysis will analyze how different people benefit from the ML model after it plateaus for a user. The online learning assesses to the practicality of the model to predict user decisions while learning user preferences in a real-world setup, contrasting all of the offline analysis so far in the paper.

This sections presents a strategy the platform can follow in the learning phase of a new user.
The key objective of the learning strategy should be to learn most of the user's privacy preferences
with minimal user involvement (prompts). 
%To better understand how to reduce prompting, while maintaining model accuracy over the status quo, we first examine how prompts affect model accuracy. This section presents an analysis of how the ML model's accuracy changes as prompts increase. Since a fully trained model requires twice as many prompts as AOFU, it is necessary to understand how the ML model behaves with fewer prompts. 
Once the model reaches adequate training, we can use model decision confidence to analyze how the ML model performs for different users and examine the tradeoff between user involvement and accuracy. We also utilize the model's confidence on decisions to present a strategy that can further reduce model error through selective permission prompting.

\subsection{Bootstrapping}
\label{sec:bootstrapping}

The \textit{bootstrapping} phase occurs when the ML model is presented with a new user about whom the model has no prior information. In this section, we analyze how the accuracy improves as we prompt the user. Since the model presented in \S\ref{sec:learning} is a single model trained with data from all users, the ML model can still predict a new user's privacy decisions by leveraging the data collected on other users' preferences.

We measured the accuracy of the ML model as if it had to predict each user's
prompt responses using a model trained using other users' data. Formally, this is called leave-one-out cross-validation, where we remove all the prompt responses from a single user. The training set contains all the prompt responses from 130 users and the test set is the prompt responses collected from the single remaining user. The model had a median accuracy of 66.6\% (56.2\% for \textit{Contextuals}, 86.4\% for \textit{Defaulters}). Although this approach does not prompt new users, it falls short of AOFU\@. This no-prompt model behaves close to random guessing for \textit{Contextuals} and significantly better for \textit{Defaulters}. Furthermore, Wijesekera et al.\ found that individuals' privacy preferences varied a lot~\cite{wijesekera2015}, suggesting that utilizing other users' decisions to predict decisions for a new user has limited effectiveness, especially for \textit{Contextuals}; some level of prompting is necessary.

%This could be due to the fact that the way \textit{Contextuals} make decisions is mostly unique from others, whereas \textit{Defaulters} tend to share how they perceive threats to a certain extent.

There are a few interesting avenues to explore when determining the optimal way to prompt the user in the learning phase. One option would be to follow the same weighted-reservoir sampling algorithm mentioned in \S\ref{sec:instrumentation}. The algorithm is weighted by the frequency of each \textit{application:permission:visibility} combination. The most frequent combination will have the highest probability of creating a permission prompt and after the given combination reaches a maximum of three prompts, the algorithm will no longer consider that combination for prompting, giving the second most frequent combination the new highest probability. Due to frequency-weighting and multiple prompts per combination, the weighted-reservoir sampling approach requires more prompts to cover a broader set of combinations. However, AOFU prompts only once per combination without frequency-weighting. This may be a useful strategy initially for a new user since it allows the platform to learn about the users' privacy preferences for a wide array of combinations with minimal user interaction.

%A more suitable approach would be to weight the probabilities by frequency and after a single prompt per combination, algorithm should move to the next combination so that with minimal number of prompts platform can learn about their privacy preferences for wide array of combinations. 

To simulate such an approach, we extend the aforementioned no-prompt model (leave-one-out validation). In the no-prompt model, there was no overlap of users in the train and test set. In the new approach, the training set includes the data from other users as well as the new user's responses to the first occurrence of each unique combination of \textit{application:permission:visibility}. The first occurrence of each unique combination simulates the AOFU policy. That is, this model is bootstrapped using data from other users and then adopts the AOFU policy to further learn the current user's preferences. The experiment was conducted using the same set of features mentioned in \S\ref{sec:feature_selection} (\textit{R2} + \textit{A1} + \textit{A2} and an SVM with a RBF kernel). The test set only contained prompt responses collected after the last AOFU prompt to ensure chronological consistency.
%The test set includes prompt responses collected after the latest AOFU prompt in the training set of the given user (this is to assure that the test set is chronologically consistent).

%Values for aggregated features for the test set is calculated based on the prompt responses for the first \textit{n} AOFU responses. The denial rate of the first \textit{n} responses and their first response for each unique combination were used to bias the aggregated features towards denying or allowing permissions in the test set. The adjustment of aggregated features based on AOFU responses resulted in a significant gained in the accuracy.

%To understand how we can incorporate AOFU to the ML model so that we can achieve higher accuracy with minimal number of prompts, we simulated a platform where the model prompts the user based on AOFU policy (i.e. rather than prompting multiple times per combination, it only prompts at the first occurrence of each new combination.). However we assumed that the ML model is already trained on data collected from other users thus in each case the training set for the model would be the data collected from other users and the prompt responses for first \textit{n} AOFU prompts of the interested user; test set is the remaining prompt responses from the given user.

\begin{figure}[t]
\centering
\includegraphics[scale=0.4]{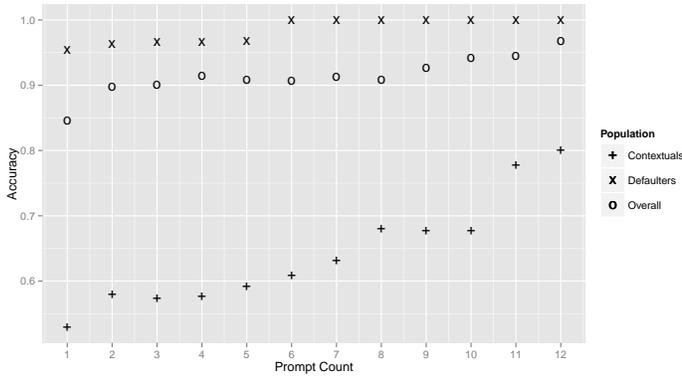}
\begin{flushleft}
\caption{How the median accuracy varies with the number of seen prompts}
\label{fig:prompt_aofu}
\end{flushleft}
\end{figure}

Figure \ref{fig:prompt_aofu} shows how accuracy changes with the varying number of AOFU prompts for \textit{Contextuals} and \textit{Defaulters}. For each of the 131 users, we ran the experiment varying the AOFU prompts from 1 to 12. We chose this upper bound because, on average, a participant saw 12 different unique \textit{application:permission} combinations during the study period---the current permission model in Android. %During the study period, if the AOFU policy was in place with \textit{application:permission:visibility}, a user would have seen a minimum of 16 prompts. 
AOFU relies on user prompts for each new combination. The proposed ML model, however, has the advantage of leveraging data collected from other users to predict a combination not seen by the user; it can significantly reduce user involvement in the learning phase. After 12 prompts, accuracy reached 96.8\% across all users.

%. Thus, AOFU needs to prompt at least 16 times before it can even make a prediction for all decisions. On the other hand, our hybrid approach does not have to prompt 16 times to predict privacy decisions across all the different combinations; this is because the model is already trained using other users' data. Hence, the hybrid approach can reach similar to or greater accuracy than AOFU with fewer prompts.

Each new user starts off with a single model 
shared by all new users and then gradually moves onto a separate model trained with AOFU prompt responses. We analyze its performance for 
\textit{Defaulters} and \textit{Contextuals} separately, finding that it 
improves accuracy while reducing user involvement in both cases, compared to the 
status quo. 

We first examine how our model performs for \textit{Defaulters}, 
53\% of our sample. Figure \ref{fig:prompt_aofu} shows that our model trained 
with AOFU permission-prompt responses outperforms AOFU from the very beginning. 
The model starts off with 96.6\% accuracy 
(before it reaches close to 100\% after 6 prompts), handily exceeding AOFU's 
93.33\%. This is a 83.3\% reduction in permission prompts compared to AOFU-AP 
(the status quo). Even with such a significant reduction in user involvement, 
the new approach cuts the prediction error rate in half.

\textit{Contextuals} needed more prompts to outperform the AOFU policy; the hybrid approach matches AOFU-AP with just 7 prompts, a 42\% reduction in prompts. With 12 permission prompts, same as needed for AOFU-AP, the new approach had reduced the error by 43\% over AOFU-AP\@(the status quo). The number of prompts needed to reach this level of accuracy in the new approach is 25\% less than what is needed for AOFU-APV\@. We also observed that as the number of prompts increased, the AUC of our predictions also similarly increased. Overall, the proposed learning strategy reduced the error by 80\% after 12 user prompts over AOFU-AP. Given, \textit{Defaulters} plateaus early in their learning cycle (after only 6 prompts), the proposed learning strategy, on average, need 9 prompts to reach it's maximum capacity which is a 25\% reduction in user involvement over AOFU-AP. 

\textit{Contextuals} have a higher need for user involvement than \textit{Defaulters}, primarily because it is easy to learn about \textit{Defaulters}, as they are more likely to be consistent with early decisions. On the other hand, \textit{Contextuals} vary their decision based on different contextual cues and require more user involvement for the model to learn the cues used by each user. Thus, it is important to find a way to differentiate between \textit{Defaulters} and \textit{Contextuals} early in the bootstrapping phase to determine which users require fewer prompts. The analysis of our hybrid approach addresses the concern of a high number of permission prompts initially for an ML approach. Over time, accuracy can always be improved with more prompts.

Our new hybrid approach of using AOFU-style permission prompts in the bootstrapping phase to train our model can achieve much higher accuracy than AOFU, with significantly fewer prompts. Having a learning strategy (use of AOFU) over random selection helped to minimize the user involvement (24 vs. 9) while significantly reducing the error rate (7.6\% vs. 3.2\%) over a random selection of the training set.

%The phase the ML model goes through until it reaches its full prediction power (i.e., accuracy of 95.73\% in the ML model and 80.53\% in AOFU) is called as the bootstrapping phase. While, more prompting generates more ground truth that will help to fine tune the ML model to accommodate specific preferences, more prompting will likely to create user habituation and user annoyance which will diminish the reliability of the ground truth and in turn the accuracy of the trained ML model. Hence, it is crucial to learn the users privacy preference with minimal user involvement.

%Based on the prompt responses, Table \ref{tbl:aofu_perf} shows that to reach its full predicting capacity, AOFU needs to prompt the user on average 12 prompts. This number varies with the number of number of unique application:permission combinations the platform has seen. For the ML model, the weighted reservoir sampler has prompted a user 32 times on average during the study period. Figure \ref{fig:prompt_acc} shows hows the accuracy of the ML model changes with the number of prompts a user has seen.

\subsection{Decision Confidence} 
\label{sec:steady_state}

In the previous section, we looked into how we can optimize the learning phase 
by merging AOFU and the ML model to reach higher accuracy with minimal user 
prompts. However, for a small set of users, more permission prompts will not 
increase accuracy, regardless of user involvement in the bootstrapping phase. 
This could be due to the fact that a portion of users in our dataset are making 
random decisions, or that the features that our ML model takes into account are 
not predictive of those users' decision processes. While we do not have the data 
to support either explanation, we examine how we can measure whether the ML 
model will perform well for a particular user and quantify how often it does 
not. We present a method to identify difficult-to-predict users and reduce permission 
prompting for those users.

While running the experiment in \S\ref{sec:bootstrapping}, we also measured how confident the ML model was for each decision it made. To measure the ML model's confidence, we record the probability for each decision; since it is a binary classification (deny or allow), the closer the probability is to 0.5, the less confident it is. We then chose a \textit{class probability threshold} above which a decision would be considered a high-confidence decision. In our analysis, we choose a class probability threshold of 0.6, since this value resulted in $>$96\% accuracy for our fully-trained model ($\approx$25 prompts per user) for high-confidence decisions, but this is a tunable threshold. Thus, in the remainder of our analysis, decisions that the ML model made with a probability of $>$0.60 were labeled as high-confidence decisions, while those made with a probability of $<$0.60 were labeled as low-confidence decisions.

Since the most accurate version of AOFU uses 12 prompts, we also evaluate the confidence of our model after 12 AOFU-style prompts. This setup is identical to the bootstrapping approach; the model we evaluate here is trained on responses from other users and the first 12 prompts chosen by AOFU\@. With this scheme, we found that 10 users (7.63\% of 131 users) had at least one decision predicted with low confidence. The remaining 92.37\% of users had all privacy decisions predicted with high confidence. Among those users whose decisions were predicted with low confidence, the proportion of low-confidence decisions on average accounted for 17.63\% (median = 16.67\%) out of all their predicted decisions. With a sensitive permission request once every 15 seconds~\cite{wijesekera2015}, prompting even for 17.63\% of predictions is not practical. Users who had low-confidence predictions had a median accuracy of 60.17\%, compared to 98\% accuracy for the remaining set of users with only high-confidence predictions. Out of the 10 users who had low-confidence predictions, there were no \textit{Defaulters}. This further supports the observation in Figure~\ref{fig:prompt_aofu} that \textit{Defaulters} require a shorter learning period.

%The difference in the accuracy highlights that the ML model will behave relatively badly for low confident predictions thus it might require more user involvement to increase the accuracy. 

%However, we did not observe any significant change among the users who had predictions with a low confidence as we increase the number of initial AOFU prompts. Thus, we believe that these set of users will continue to have predictions made with a lower confidence regardless of the user involvement and if the platform continues to prompt the user with the aim of reducing the low confident prediction portion, it will only habituate the user and reduce the reliability of the ground truth which could in turn worsen the future predictions.

In a real-world scenario, after the platform (ML model) prompts the user for the first 12 AOFU prompts, the platform can measure the confidence of predicting unlabeled data (sensitive permission requests for which the platform did not prompt the user). If the proportion of low-confidence predictions is below some threshold, the ML model can be deemed to have successfully learned user privacy preferences and the platform should keep on using the regular permission-prompting strategy. Otherwise, the platform may choose to limit prompts (i.e., two per unique \textit{application:permission:visibility} combination). It should also be noted that rather than having a fixed number of prompts (e.g., 12) to measure the low-confidence proportion, the platform can keep track of the low-confidence proportion as it prompts the user according to any heuristic (i.e., unique combinations). If the proportion does not decrease with the number of prompts, we can infer that the ML model is not learning user preferences effectively or the user is making random decisions, indicating that limiting prompts and accepting lower accuracy could be a better option for that specific user, to avoid excessive prompting. However, depending on which group the user is in (\textit{Contextual} or \textit{Defaulter}), the point at which the platform could make the decision to continue or limit prompting could change. In general, the platform should be able to reach this deciding point relatively quickly for \textit{Defaulters}. 

Among participants with no low-confidence predictions, we had a median error rate of 2\% (using the new hybrid approach after just 12 AOFU prompts); for the same set of users, AOFU reached a median error rate of 13.3\%. However, using AOFU, a user in that set would have needed an average of 15.11 prompts to reach that accuracy. Using the ML model, a user would need just 9 prompts on average (\textit{Defaulters} require far fewer prompts, dropping the average); the model only requires 60\% of the prompts that AOFU requires. Even with far fewer prompts in the learning phase, the ML model achieves a 84.61\% reduction in error rate as compared to AOFU\@.

While our model may not perform well for all users, it does seem to work quite well for the majority of users (92.37\% of our sample). We provide a way of quickly identifying users for whom our system does not perform well, and propose limiting prompts to avoid excessive user burden for those users, at the cost of reduced efficacy. In the worst case, we could simply employ the AOFU model for users our system does not work well for, resulting in a multifaceted approach that is at least as good as the status quo for all users.

\subsection{Online Model} 
\label{sec:online_model}
Our proposed system relies on training models on a trusted server, sending it to client phones (i.e., as a weight vector), and having phones make classifications. By utilizing an online learning model, we can train models incrementally as users respond to prompts over time. There are two key advantages to this: (i) this model adapts to changing user preferences over time; (ii) training models on multiple users' data allows more labeled data points for training.

Our scheme requires two components: a feature extraction and storage mechanism on the phone (a small extension to our existing instrumentation) and a machine learning pipeline on a trusted server. The phone sends feature vectors to the server every few prompts, and the server responds with a weight vector representing the newly trained classifier. To bootstrap the process, the server's models can be initialized with a model trained on a few hundred users, such as our single model across all users. Since each user contributes data points over time, the online model adapts to changing privacy preferences even if they conflict with previous data. When using this scheme, each model takes less than 10\,KB to store. With our current model, each feature and weight vector are at most 3\,KB each, resulting in at most 6\,KB of data transfer per day.

To evaluate the accuracy of our online model, we trained a classifier using stochastic gradient descent (SGD) with five-fold cross validation on our 4,224-point data set. This served as the bootstrapping phase. We then simulated receiving the remaining data one-at-a-time in timestamp order. Any features that changed with time (e.g., running averages for aggregate features, event counts) were computed with each incoming data point, creating a snapshot of features as the phone would see it. We then tested accuracy on the chronologically last 20\% of our dataset. Our SGD classifier had 93.8\% accuracy (AUC=0.929). We attribute the drop in accuracy (compared to our offline model) to the fact that running averages take multiple data points to reach steady-state, causing some earlier predictions to be incorrect.

A natural concern with a trusted server is compromise. To address this concern, we do not send any personally-identifiable data to the server. Furthermore, features sent to the server have been \textit{scaled}; they are reported in standard deviations from the mean, not in raw values.

\section{Contextual Integrity}
\label{sec:contextual-integrity}

%Contextual integrity provides a conceptual framework to better understand how
%users make privacy decisions; we utilize the formalized model of contextual
%%integrity introduced in \cite{Barth2006} as a framework to view existing and
%our Android permission systems in. In the formal model, they model parties as
%communicating agents ($P$) knowing information about other agents, represented
%as attributes ($T$). Agents play roles ($R$) in contexts ($C$), and agents can
%play multiple roles. For example, an agent can be a game application, and have
%the role of a game provider in an entertainment context. Knowledge transfer
%happens when information is communicated between agents. Communications can
%only occur when they follow the norms of context; the relationship between the
%agent sending the information and the role of agent receiving it must follow
%these norms too. For example, it follows norms for a patient to provide
%medical information to her doctor.

Contextual integrity is a conceptual framework that helps explain
why most permission models fail to protect user privacy---they often do not
take the context surrounding
privacy decisions into account. In addressing this issue, we propose an ML
model that infers when context has changed. That is, if the system knows that a user is comfortable sharing data with a particular application under one set of circumstances, it should not bother her with a permission request when the same application requests access to the same data under similar circumstances in the future. However, it should behave differently when those circumstances have changed. We believe that this is an important first step
towards operationalizing the notion of \emph{contextual integrity}. In this
section, we explain the observations that we made in \S\ref{sec:accuracy} within the context of the contextual integrity framework proposed by Barth et al.~\cite{Barth2006}.

Contextual integrity provides a conceptual framework to better understand how
users make privacy decisions; we use Barth et al.'s formalized
model~\cite{Barth2006} as a framework in which to view Android permission
models. Barth et al.~model parties as communicating agents ($P$) knowing
information represented as attributes ($T$). A knowledge state $\kappa$ is
defined as a subset of $P\times P\times T$\@. We use $\kappa = (p,q,t)$ to mean
that agent $p$ knows attribute $t$ of agent $q$. Agents play roles ($R$) in
contexts ($C$).

For example, an agent can be a game application, and have the role of a game
provider in an entertainment context. Knowledge transfer happens when
information is communicated between agents; all communications can be
represented through a series of traces $(\kappa, (p,r), a)$, which are
combinations of a knowledge state $\kappa$, a role state $(p,r) $, and a
communication action $a$ (information sent). The role an agent plays in a
given context helps determine whether an information flow is acceptable for a
user. Communications can only occur when they follow the norms of context; the
relationship between the agent sending the information and the role of the agent
($(p,r)$) receiving it must follow these norms, too.

With the Android permission model, the same framework can be applied. 
Both the user and the third-party application are communicating agents, and
the information to be transferred is the sensitive data requested by the
application.  When a third-party application requests permission to access a
guarded resource (e.g., location), knowledge of the guarded resource
is transferred from the one agent (i.e., the user/platform) to another agent
(i.e., the third-party application). The
extent to which a user expects a given request
 depends not on the agent (the application requesting the data), but on the 
role that agent is playing in that context. This explains why the application as a 
feature itself (i.e., application name) was not predictive in our models:
this feature does not represent the role
when determining whether it is unexpected. While it is hard, from the platform,
to determine the exact role an application is playing,
the visibility of the application 
hints at its role. For instance, when the user is using Google Maps to navigate, it 
is playing a different role from when Google Maps is running in the background without 
the user's knowledge. We believe that this is the reason why the visibility of the requesting 
application is significant: it helps the user to infer the role played
by the application requesting the permission.

The user expects applications in certain roles to access
resources depending on the context in which the request is made. We believe that
the foreground application sets this context. Thus a combination of the role
and the context decides whether an information flow is expected to occur or
not. Automatically inferring the exact context of a request (e.g., how data will be used, whether it will be shared with others, etc.) is likely an intractable problem. However, for our purposes, it is possible that we need to only infer when context has {\it changed}, or rather, when data is being requested in a context that is no longer acceptable to the user. Based on our data, we believe that features based on 
foreground application and visibility are most useful for this purpose, amongst the features that we collected during this study.

We now combine all of this into a concrete example within the contextual
integrity framework: If a user is using Google Maps to reach a destination, the
application can play the role of a navigator in a geolocation context, whereby
the user feels comfortable sharing her location. In contrast, if the same
application requests location while running as a service invisible to the user,
the user may not want to give this service the same information. Background applications play the role of ``passive listeners'' in
most contexts; this role as perceived by the user may be why background
applications are likelier to violate privacy expectations and
consequently be denied information by users.

AOFU primarily focuses on controlling access through rules for \textit{application:permission}
combinations. Thus, AOFU neglects the role played by the application
(visibility) and relies purely on the agent (the application)
and the information subject (permission type). This explains why
AOFU is wrong in nearly one-fifth of cases. Based on
Table~\ref{tbl:aofu_perf}, both AOFU-VA (possibly identifying the role
played by the application) and
AOFU-A$_F$PV (possibly identifying the current context because of the
current foreground application-A$_F$) have higher accuracy
than the other AOFU combinations. However, as the
contextual integrity framework suggests, the permission model has to take both the
role and the current context into account before making an accurate decision.
AOFU (and other models that only consider the application name and permission type) only makes it possible to consider a single aspect, a limitation that does
not apply to our model.

%SE: I commented this out below, I found it a bit confusing, and it doesn't seem to add much.
%The aggregate feature that combines visibility with the foreground application
%helps to capture the role and the context to a greater extent, which in turn
%reduces errors significantly. We only encompass role probabilistically: the
%aggregate feature and the denial rate roughly estimate the probability that an
%application plays a role that the user would grant the information to.
%To better understand roles, our instrumentation would need to be extended into the application space. Overall, this framework helps us formalize that by encapsulating more context to preserve contextual integrity, our model is able to achieve better accuracy.

While the data presented in this work suggest the importance of capturing
context to more accurately protect user privacy, more work is needed along these
lines to fully understand how people use context to make decisions in the Android permission model. Crucially, research to better define context.
Nevertheless, we believe we contribute a significant initial step towards
applying contextual integrity to improve smartphone privacy by dynamically regulating permissions.

\section{Discussion}
\label{sec:disc}

The primary goal of this research was to improve the accuracy of the Android permission
system so that it more correctly aligns with user privacy preferences. We began with four
hypotheses: (i) that the currently deployed AOFU policy frequently
violates user privacy; (ii) that the contextual information it ignores is useful; (iii) that a
ML-based classifier can account for this contextual information and thus improve on the status quo; and
(iv) that passively observable behavioral traits can be used to infer privacy preferences.

To test these hypotheses, we performed the first large-scale study on the
effectiveness of AOFU permission systems in the wild, which showed that hypotheses (i)
and (ii) hold. We further built an ML classifier that took user permission
decisions along with observations of user behaviors and the context surrounding those decisions to show that (iii) and (iv) hold.
Our results show that existing systems have
significant room for improvement, and other permission-granting systems may benefit from applying our results.

\subsection{Limitations of Permission Models}

Our field study confirms that
users care about their privacy and are wary of permission
requests that violate their expectations.
We observed that 95\% of participants chose to block at least one
permission request; in fact, the average denial rate was 60\%---a staggering
amount given that Android's earlier AOI model permits all
permission requests once an application is installed. This denial rate implies
that AOI correctly regulates permission requests only one in four times.

While AOFU improves over the AOI model, it still violates
user privacy around one in seven times, as users deviate from their initial responses to
permission requests. This amount is significant because of
the high frequency of sensitive permission requests: a 15\% error rate
yields thousands
of privacy violations for each user \emph{per day}.
It further shows that AOFU's correctness assumption---that users
make binary decisions based only on the \emph{application:permission}
combination---is incorrect. Users take a
richer space of information into account when making decisions about permission requests.

\subsection{Our ML-Based Model}
We show that ML techniques are effective
at learning from both the user's previous decisions and the
current environmental context in order to predict whether to grant permissions
on the user's behalf. In fact, our techniques achieve better results than the
methods currently deployed on millions of phones worldwide---while imposing 
significantly less user burden.

Our work incorporates elements of the environmental context into a machine-learning
model. This better approximates user decisions by finding factors relevant for
users that are not encapsulated by the AOFU model.
In fact, our ML model reduces the errors made by the AOFU
model by 75\%. Our ML model's 97\% accuracy is a substantial improvement over
AOFU's 85\% and AOI's 25\%; the latter two of which comprise the {\it status
quo} in the Android ecosystem.

Our research shows that many users make neither random nor fixed decisions:
the environmental context plays a significant role in user decision-making.
Automatically detecting the precise context surrounding a request for sensitive data is an incredibly difficult problem (e.g., inferring {\it how} data will be used), and is potentially intractable. However, to better support user privacy, that problem does not need to be solved; instead, we show that systems can be improved by using environmental data to infer when context has {\it changed}. We found that the most predictive factors in the environmental context
were whether the application
requesting the permission is visible, and if not, what foreground application
actually was visible. These are
both strong contextual cues used by users, insofar as they allowed us to better predict changes in context. Our results show that ML techniques
have great potential in improving user privacy, by allowing us to infer when context has changed, and therefore when users would want data requests to be brought to their attention.

\subsection{Reducing the User Burden}
Our work is also novel in using passively observable data to infer privacy
decisions: we show that we can predict a user's preferences without \emph{any}
permission prompts.  Our model trained solely on behavioral traits yields a
three-fold improvement over AOI; for \textit{Defaulters}---who account for
53\% of our sample---it was as accurate
as AOFU-AP\@. These results demonstrate that we can match the status quo without
\textit{any} active user involvement (i.e., the need for obtrusive prompts). These results imply that learning privacy
preferences may be done entirely passively, which, to our knowledge, has not yet been attempted in this domain. Our behavioral feature set provides a promising new direction to guide research in creating permission models that minimize user burden.

The ML model trained with contextual data and past decisions also significantly reduced
the user burden while achieving higher accuracy than AOFU. The model yielded an 81\% reduction
in prediction errors while reducing user involvement by 25\%. The significance of this observation
is that by reducing the risk of habituation, it increases reliability when user input is needed.

\subsection{User- and Permission-Tailored Models}
Our ML-based model incorporates data from all users into a single predictive
model. It may be the case, however, that a collection of models tailored to
particular types of users outperforms our general-purpose model---provided
that the correct model is used for the particular user and permission.
To determine if this is true, we
clustered users into groups based first on their behavioral features, and
then their denial rate, to see if we could build superior cluster-tailored ML
models. Having data for only 131 users, however,
resulted in clusters too small to carry out an effective analysis. We note that we also created a separate model for each sensitive permission
type, using data only for that permission. Our experiments determined, however,
that these models were no better (and often worse) than our general model. It is possible that such tailored models may be more useful when our system is implemented at scale.

%We hope that
%future work builds on our results and increases the data available for analysis.
%In particular, if our model is integrated into future Android devices, a
%trusted third-party server could aggregate permission-prompt data
%from the large user base. This data can be used to train more sophisticated
%permission systems with clustering that may outperform the one we have developed.

\subsection{Attacking the ML Model}

%\section{Attacking the ML Model}
%\label{sec:attack_ml}
%
Attacking the ML model to get access to users' data without
prompting is a legitimate concern~\cite{barreno2006can}.
There are multiple ways an adversary can influence the proposed permission
model: (i) imposing an adversarial ML environment~\cite{lowd2005adversarial};
(ii) polluting
the training set to bias the model to accept permissions; and (iii) manipulating input features in order to get access without
user notification. We assume in this work that the platform is not compromised; a compromised platform will degrade any permission model's ability to protect resources.

A thorough analysis on this topic is outside of our scope. Despite that,
we looked at the possibility of manipulating
features to get access to resources without user consent.
None of the behavioral features used in the model can be influenced, 
since that would require compromising the platform.
An adversary can control the runtime features for a given permission
request by specifically choosing when to request the permission.
We generated feature vectors that encompassed every adversary-controlled 
value and combination from our dataset, and tested them on our model.
We did not find any conclusive evidence that the adversary can exploit the
ML model by manipulating the input features to get access to resources
without user consent.

As this is not a comprehensive analysis on attack vectors, it is possible that
there exists a scenario where the adversary is able to access sensitive
resources without prompting the user first. Our preliminary analysis suggests
that they may be non-trivial, but more work is needed to study and prevent such
attacks. In particular, to protect against adversarial ML techniques and
formally examining feature brittleness.

\subsection{Experimental Caveat}

We repeat a caveat about our experimental data: users were free to
deny permissions without any consequences.
%denying a legitimately-needed permission did not result in loss of functionality.
We explicitly informed participants
in our study that their decisions to deny permission requests would have no impact on
the actual behavior of their applications. This is important to note because if
an application is denied a permission, it may exhibit undefined behavior or lose
important functionality. If these consequences are imposed on users, they may
decide that the functionality is more important than their privacy decision. 

%Similarly, the loss of functionality may demonstrate the necessity of allowing certain permissions that are otherwise unclear.

If we actually denied permissions, users' decisions may skew towards
a decreased denial rate. The denial rates in our experiments therefore represent the
actual privacy preferences of users and their
{\it expectations} of reasonable application behavior---not the result of choosing
between application functionality and privacy. We believe that
how people react when choosing between functionality and privacy
preferences is an important research question beyond the scope of this paper.

We believe that there are important unanswered questions about how to solve the technical
hurdles surrounding enforcing restrictive preferences with minimal usability issues. In
fact, researchers have noted that many applications crash when permissions
are denied~\cite{revdroid16}. As a first step towards building a platform
that does not force users to choose between their privacy preferences and required
functionality, we must develop an environment where
permissions appear---to the application---to be allowed, but in reality only
spurious or artificial data is provided. 

%It is possible, for instance, that those categorized as \textit{Defaulters} are an artifact of our experiment, as denying all permissions had no consequences. Yet, limiting our analysis to \textit{Contextuals} does not limit our claims.

We leave as future work the replication of this experiment with consequences
for denied application permissions. 
We expect that a considerable
portion of the default-deny contingent would become more selective with their
privacy preferences in the presence of actual permission denial. Such a
change, however, will not limit this contribution, since our proposed model
was effective in guarding resources of the users who are selective in their
decision making---the proposed classifier was able to reduce the error rate of
Contextuals by 44\%.

%Note that the instrumentation of the Android
%platform to seamlessly provide this is non-trivial because many applications are not
%programmed to correctly handle denied permissions. This is despite modern
%Android already empowering users to deny permissions on their first use.  Consequently,  Such an experiment should provide the most accurate
%user permission data ever collected, and 

\subsection{Types of Users}
We presented a categorization of users based on the significance that the
application's visibility played towards their individual
privacy decisions. We believe that in an actual permission denial setting, the distribution
will be different from what was observed in our study. Our
categorization's significance, however, motivates a
deeper analysis on understanding the factors
that divide \textit{Contextuals} and \textit{Defaulters}. We believe that
visibility is an important factor in this division but there may be others that
are more significant. More work needs to be done to explore
how \textit{Contextuals} make decisions and which behaviors correlate with their decisions.

\subsection{User Interface Panel}
Any model that predicts user decisions has the risk of making
incorrect predictions. Making predictions on a user's behalf, however,
is necessary
because permissions are requested by applications with too high a frequency for
manual examination.
%Frequent prompting further contributes to user habituation
%and warning fatigue: more-frequent prompts yield less-thoughtful responses.
Thus, platforms need to make these predictions and should strive to be as
accurate as possible. While we do not expect any system to be
able to obtain perfect accuracy, we do expect that our 97\% accuracy can be
improved upon.

% to->LYNN: add UI interface, user interaction, feedbackloop etc. here.
% UI also to allow users to harden security against an app, e.g., because of a
% read article in news, 

% from -> LYNN: I added some more information regarding the UI and interaction,
% I'm not exactly sure what you mean by feedback loop though?

%Users face many decisions when interacting with permissions. The complexity of
%these interactions may increase the burden on users. By simplifying and
%clarifying the interactions between user and permissions, we believe users
%could gain a better understanding and therefore make more informed decisions.
%This would also be left to future work. By creating an application for the
%users that is linked to their permission data, we allow users to view and
%adjust the decisions made by the machine learning model. 

One plausible way of improving the accuracy of the permission model is to
empower the user to review and make changes on how the ML model makes decisions
through a user feedback panel.  A
major benefit is that users would be able to go back and review the decisions
made by the ML model. It would also allow users to adjust these decisions
according to their preferences, thereby correcting errors. This gives users recourse to
correct undesirable decisions. The UI panel could also be used to reduce the usability issues and
functionality loss stemming from permission denial.  The panel should help the
user figure out which rule incurred the functionality loss and to change it
accordingly. A user may also use this to adjust their settings as their privacy preferences evolve over time.

%or due to external sources, such as reading an article in the news.

%In the event
%of unexpected functionality loss due to a denial, the user should be able
%to fix it through a UI panel.

%Furthermore, users
%would be aware of what functionality loss was incurred due to which
%permissions, and adjust their decision to best fit their needs. For example, if
%denying a permission results in complete loss of functionality for a particular
%application, the user may adjust his or her decision. 

%The interface would also provide the user information on his or her own
%decisions, and a general overview of which applications requested which
%permissions.  Eventually, the panel may be able to incorporate information
%regarding \textit{why} an application needs a set of permissions being
%requested, giving users more context into an applications permission requests.
%Knowing this information, a user would be able to better learn and decide what
%he or she would prefer and make more accurate and desirable decisions. Finally,
%the panel should also preserve the contextuality behind each rule so that users
%can still modify it without losing the contextual aspect of their previous decision.
%It is important that this interface be straightforward and easy to understand.
%Our target audience is any potential Android user and
%therefore it must be as presented in a manner that is understandable by
%virtually any population of users. 

\subsection{Conclusions}

We have shown a number of important results. Users care about their privacy: 
they deny a significant number of requests to access sensitive data. Existing
permission models for Android phones still result in significant privacy
violations. Users may allow permissions some times, while denying them at
others, which means that there are more factors that go into the
decision-making process than simply the application name and the permission
type. We collected real-world data from 131 users and found that application
visibility and the current foreground application were important factors in user
decisions. We used the data we collected to build a
machine-learning model to make automatic permission decisions. One of our models,
without any user prompting, had a comparable error rate to what AOFU has, and another of our
models reduced the number of errors by 81\% with even less prompting.


\section*{Acknowledgments}

This research was supported by the United States Department of Homeland Security's Science and Technology Directorate under contract FA8750-16-C-0140, the Center for Long-Term Cybersecurity (CLTC) at UC Berkeley, and the National Science Foundation under grant CNS-1318680. The content of this document does not necessarily reflect the position or the policy of the U.S. Government and no official endorsement should be inferred.



%\bibliographystyle{IEEEtranS}
%\bibliography{IEEEabrv,sigproc}

% Generated by IEEEtranS.bst, version: 1.12 (2007/01/11)

%\begin{thebibliography}{10}
%\end{thebibliography}
\newpage
\appendices

\section{Information Gain of Features}
\label{app:infogain}

\begin{table}[h]
\centering
\begin{tabular}{|l|l|l|l|}
\hline
 & Contextuals & Defaulters & Overall \\ \hline
A1 & 0.4839 & 0.6444 & 0.5717 \\ \hline
A2 & 0.4558 & 0.6395 & 0.5605 \\ \hline
Permission & 0.0040 & 0.0038 & 0.0050 \\ \hline
Time & 0.0487 & 0.1391 & 0.0130 \\ \hline
Visibility & 0.0015 & 0.0007 & 0.0010 \\ \hline
\end{tabular}
\caption{Feature Importance of Contextual Features}
\label{tbl:inforgain}
\end{table}

\section{Information Gain of Behavioral Features}
\label{app:behaveinforgain}

\begin{table}[h]
\centering
\begin{tabular}{|l|l|}
\hline
\textbf{Feature} & \textbf{Importance} \\ \hline
Amount of time spent on audio calls & 0.327647825 \\ \hline
Frequency of audio calls & 0.321291184 \\ \hline
\begin{tabular}[c]{@{}l@{}}Proportion of times screen was timed out\\ instead of pressing the lock button\end{tabular} & 0.317631096 \\ \hline
\begin{tabular}[c]{@{}l@{}}Number of times PIN was used to\\ unlock the screen.\end{tabular} & 0.305287288 \\ \hline
Number of screen unlock attempts & 0.299564131 \\ \hline
Amount of time spent unlocking the screen & 0.29930659 \\ \hline
Proportion of time spent on loud mode & 0.163166296 \\ \hline
Proportion of time spent on silent mode & 0.138469725 \\ \hline
\begin{tabular}[c]{@{}l@{}}Number of times a website is loaded to\\ the Chrome browser\end{tabular} & 0.094996437 \\ \hline
\begin{tabular}[c]{@{}l@{}}Out of all visited websites, the proportion\\ of HTTPS-secured websites.\end{tabular} & 0.071096898 \\ \hline
\begin{tabular}[c]{@{}l@{}}Number of times Password was used to\\ unlock the screen\end{tabular} & 0.067999523 \\ \hline
\begin{tabular}[c]{@{}l@{}}Proportion of websites requested location\\ through Chrome\end{tabular} & 0.028404167 \\ \hline
Time & 0.019799623 \\ \hline
The number of downloads through Chrome & 0.014619351 \\ \hline
Permission & 0.001461635 \\ \hline
Visibility & 0.000162166 \\ \hline
\end{tabular}
\caption{Feature Importance of Behavioral Features}
\label{tbl:behaveinforgain}
\end{table}

\end{document}